\documentclass[a4paper,aps,dvips,floatfix,groupedaddress,prl,showpacs,twocolumn]{revtex4}

\usepackage{amsmath, amssymb}
\usepackage[T1]{fontenc}
\usepackage{graphicx, epstopdf}
\usepackage[latin1]{inputenc}
\usepackage{latexsym}
\usepackage{mathrsfs}
\usepackage{textcomp}
\usepackage{times} 
\usepackage[hyperfootnotes=false]{hyperref}

\newcommand{\degree}{\ensuremath{\text{\textdegree}}}
\newcommand{\Es}{\ensuremath{\text{E}_\text{stat}}}
\newcommand{\Ey}{\ensuremath{\text{E}_\text{YAG}}}

\begin{document}

\title{Laser-induced alignment and orientation of quantum-state-selected large molecules}

\author{Lotte Holmegaard$^1$}%
\author{Jens H. Nielsen$^2$}%
\author{Iftach Nevo$^{1}$}%
\author{Henrik Stapelfeldt$^{1,3}$}%
\email{henriks@chem.au.dk}%
\affiliation{$^1$\,Department of Chemistry, University of Aarhus, DK-8000 Aarhus C, Denmark \\
   $^2$\,Department of Physics and Astronomy, University of Aarhus, DK-8000 Aarhus C, Denmark \\
   $^3$\,Interdisciplinary Nanoscience Center (iNANO), University of Aarhus, DK-8000 Aarhus C,
   Denmark}

\author{Frank Filsinger}%
\author{Jochen K\"upper}%
\email{jochen@fhi-berlin.mpg.de}%
\author{Gerard Meijer}%
\affiliation{Fritz-Haber-Institut der Max-Planck-Gesellschaft, Faradayweg 4-6, 14195 Berlin,
   Germany}

\date{\today}

\begin{abstract}\noindent%
   A strong inhomogeneous static electric field is used to spatially disperse a supersonic beam of
   polar molecules, according to their quantum state. We show that the molecules residing in the
   lowest-lying rotational states can be selected and used as targets for further experiments. As an
   illustration, we demonstrate an unprecedented degree of laser-induced 1D alignment
   $\left(\langle\cos^2\theta_{2D}\rangle=0.97\right)$ and strong orientation of state-selected
   iodobenzene molecules. This method should enable experiments on pure samples of polar molecules
   in their rotational ground state, offering new opportunities in molecular science.
\end{abstract}

\maketitle

\noindent%
For many applications in physics and chemistry an ensemble of molecules all in the rotational ground
state is desirable. Such targets would provide unique possibilities, for example, for manipulating
the external degrees of freedom with static electric fields~\cite{loesch:1990:jcp,
   friedrich:1991:nature} or optical fields~\cite{fh:1995:prl, kumarappan:2006:jcp}, or
both~\cite{friedrich:1999:jcp, friedrich:1999:jpca}. A supersonic expansion of molecules in an inert
atomic carrier gas can -- to some degree -- provide such a desired molecular target: At least in the
case of small molecules (consisting of just a few atoms) the low rotational temperatures that are
obtainable result in the population of only a few rotational states. For larger polyatomic systems
rotational cooling down to or even below 1 K still leaves the molecular ensemble distributed over a
considerable range of rotational states, thereby often masking quantum state specific effects. State
selection of large molecules can be performed using inhomogeneous electric or magnetic fields. For
instance, static fields have been applied to deflect polypeptides~\cite{Broyer:ps:2007}, whereas
dynamic focusing schemes~\cite{Auerbach:JCP45:2160} have recently been used in the deceleration and
state-selection of benzonitrile~\cite{Wohlfart:PRA77:031404} and in the selection of structural
isomers of 3-aminophenol~\cite{Filsinger:prl:2008}.

Here we exploit that the lowest lying rotational states, and, eventually, the rotational ground
state, of a polar molecule can be spatially isolated by deflecting a molecular beam with a strong
inhomogeneous static electric field. The state selected molecules are used for laser induced
alignment~\cite{stapelfeldt:2003:rmp, kumarappan:2006:jcp} and mixed-field orientation
experiments~\cite{Buck:Farnik:rev:2006, minemoto:2003:jcp}. The state selection leads to strong
enhancement in the degree of orientation and alignment of iodobenzene molecules compared to that
achieved when no deflection is used. For small molecules, state-selection can be achieved using a
hexapole focuser, and it has recently been suggested that this can be used for improved alignment
and orientation experiments~\cite{Gijsbertsen:PRL99:213003}. The method demonstrated here will apply
broadly to both small and large polar molecules.

The possibility to deflect polar molecules in a molecular beam with an electric field was first
described by Kallmann and Reiche in 1921~\cite{Kallmann:ZP:1921}\footnote{Interesting enough these
   studies were performed at the \emph{Kaiser Wilhelm Institut f\"ur physikalische Chemie und
      Elektrochemie in Berlin}, the precessor of the Fritz Haber Institute.} and experimentally
demonstrated by Wrede in 1927~\cite{Wrede:ZP:1927}. As early as 1926, Stern suggested that the
technique could be used for the quantum state separation of small diatomic molecules at low
temperatures~\cite{Stern:ZP:1927}. In 1939 Rabi introduced the molecular beam resonance method, by
using two deflection elements of oppositely directed gradients in succession, to study the quantum
structure of atoms and molecules~\cite{Rabi:PR55:526}. Molecular beam deflection by static
inhomogeneous electric fields has also been used extensively as a tool to determine dipole moments
and polarizabilities of molecular systems ranging from diatomics over clusters to large
biomolecules~\cite{Broyer:ps:2007}. The present work builds on Stern's original proposal and
exploits that the force experienced by a molecule from the inhomogeneous field depends on the
rotational quantum state. In particular, the ground state has the largest Stark shift. Molecules
residing in this state are deflected most and can, therefore, be spatially separated from molecules
in other states. Our goal is to isolate and use rotational ground state molecules (or at least
molecules in the few lowest lying states) as targets for various experiments. It is crucial that the
population of ground state molecules in the molecular beam is initially as large as possible since
the deflection does not change the initial state distribution but merely disperses it. Therefore,
the rotational temperature of the molecular beam is made as low as possible using a high-pressure
supersonic expansion.

\begin{figure}
   \includegraphics[width=\linewidth]{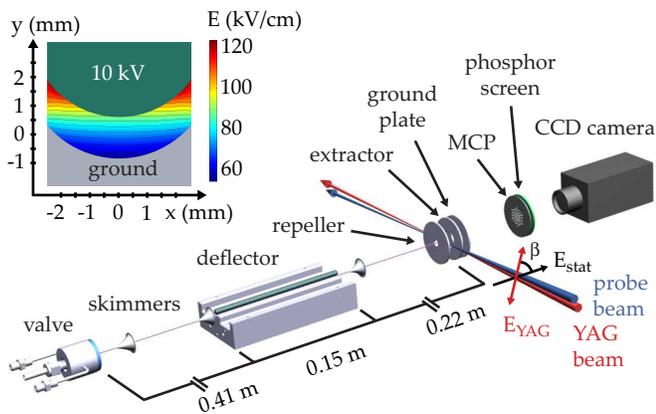}
   \caption{(Color online) Scheme of the experimental setup. In the inset, a cut through the
      deflector is shown, and a contour-plot of the electric field strength is given.}%
   \label{fig:setup}
\end{figure}

A schematic of the experimental setup is shown in Fig.~\ref{fig:setup}. About 1~mbar of iodobenzene
is expanded in 90~bar of helium into vacuum using a pulsed valve~\cite{even:2003:jcp} to produce a
molecular beam with a rotational temperature of $\sim$~1~K. After passing two 1-mm-diameter skimmers
the molecular beam enters a 15-cm-long electrostatic beam deflector. The deflector consists of a
trough (at ground potential) with an inner radius of curvature of 3.2~mm and a rod (at high voltage)
with a radius of 3.0~mm. The vertical separation of the two electrodes accross the molecular beam
axis is 1.4~mm. This electrode geometry creates a two-wire field with a nearly constant gradient
over a large area around the molecular beam axis~\cite{Ramsey:MolecularBeams:1956}.
In the inset of Fig.~\ref{fig:setup} a cross sectional view of the deflector with the created
electric field is given. The deflector is mounted such that the deflection occurs vertically, and
molecules in high-field-seeking states are deflected upwards. After passing the deflector, the
molecular beam enters the target/detection area through a 1.5-mm-diameter skimmer, where it is
crossed by one or two focused laser beams. One laser beam, consisting of 25-fs-long pulses (800~nm,
beam-waist of $\omega_0=21$~\textmu{}m), is used to probe the molecules. In the first part of the
experiment, this laser is used to characterize the deflection by determining the density at a given
height in the molecular beam via photoionization. This laser is also used for Coulomb exploding the
molecules to enable the determination of their alignment and orientation. The second laser beam,
consisting of 10-ns-long pulses from a Nd:YAG laser (1064~nm, $\omega_0=36$~\textmu{}m), is used to
align and orient the molecules. For these experiments, the probe pulse is electronically
synchronized to the peak of the YAG pulse. Ions produced by the probe pulses are accelerated, in a
velocity focusing geometry, towards a microchannel plate (MCP) detector backed by a phosphor screen.
The 2D ion images are recorded with a CCD camera. The experiments are conducted at 20~Hz, limited by
the repetition rate of the YAG laser.

\begin{figure}
   \includegraphics[width=\linewidth]{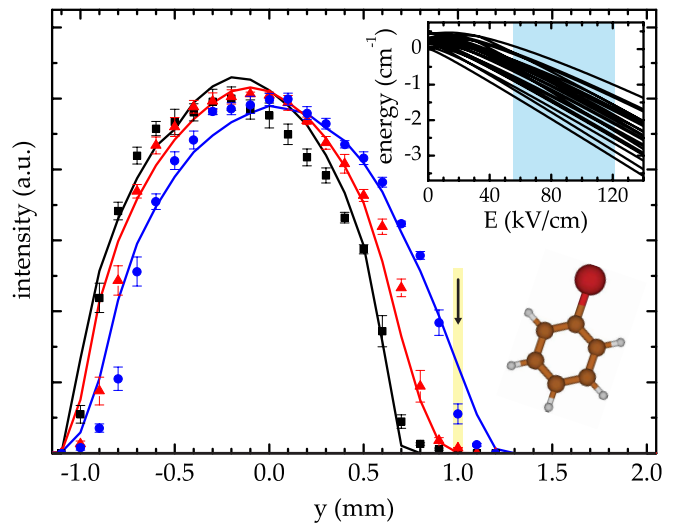}
   \caption{(Color online) The vertical profile of the molecular beam measured by recording the
      laser induced I$^+$ signal (see text). The experimental data are shown by (black) squares
      (deflector off), (red) triangles (5 kV) and (blue) circles (10 kV), together with the
      corresponding simulated profiles. In the inset the energies of the lowest quantum states of
      iodobenzene are shown as a function of the electric field strength. The shaded area represents
      the range of field strengths present in the detector for a voltage of 10 kV.}%
   \label{fig:deflection_IB}
\end{figure}

To demonstrate the effect of the deflector we measure the vertical intensity profile of the
molecular beam. This is done by recording the signal of I$^+$ ions, created by ionization with a
circularly polarized probe pulse, as a function of the vertical position of the probe laser focus,
see Fig.~\ref{fig:deflection_IB}.
When the deflector is turned off, the molecular beam extends over $\sim$~1.5 mm, mainly determined
by the diameter of the skimmer before the target area. When the deflector is turned on, the
molecular beam profiles broaden and shift upwards. The broadening and the shift become more
pronounced as the voltage on the deflector is increased from 5 kV to 10 kV. In order to simulate
these beam profiles, we first calculate the Stark curves for all rotational states from
spectroscopic constants~\cite{Dorosh:JMolSpec246:228}. The Stark energies for the lowest lying
rotational states are displayed in the inset in Fig.~\ref{fig:deflection_IB}. Then we perform
trajectory calculations for molecular packets of individual rotational states, which yield
single-quantum-state deflection profiles. These individual profiles are averaged according to the
populations for various rotational temperatures. A comparison of calculated and experimental
profiles yields a rotational temperature of the molecular beam of 1.0~K. The resulting calculated
profiles are also shown in Fig.~\ref{fig:deflection_IB} and agree well with the observations. The
calculations indicate that the most deflected molecules are indeed those initially residing in the
lowest lying rotational quantum states. In the measurements described below, experiments are
conducted on these quantum-state selected molecules simply by positioning the laser focus close to
the upper cut-off region in the 10 kV profile, as indicated by an arrow in
Fig.~\ref{fig:deflection_IB}. Similarly, experiments can be conducted on the least deflected
molecules, corresponding to molecules in high lying rotational states, by moving the lens position
to the lower part of the beam profile.

\begin{figure}
   \begin{minipage}{0.19\linewidth}
      \vspace*{0.3cm}
      \includegraphics[width=\textwidth]{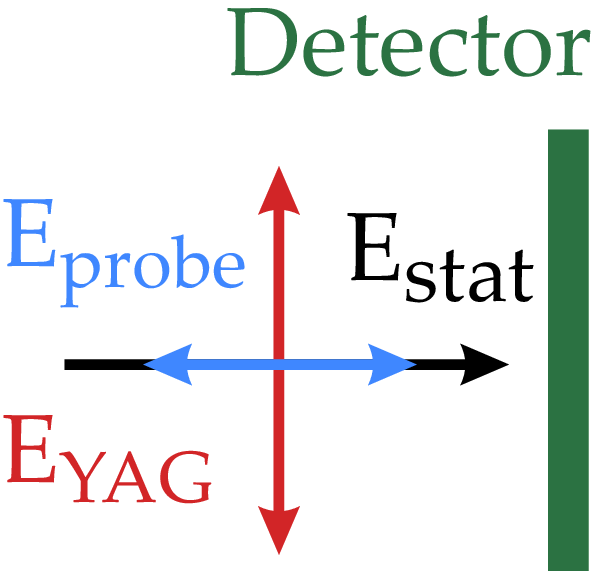}
      \vfill{\phantom{a}} \vspace{0.9cm}
      \includegraphics[width=\textwidth]{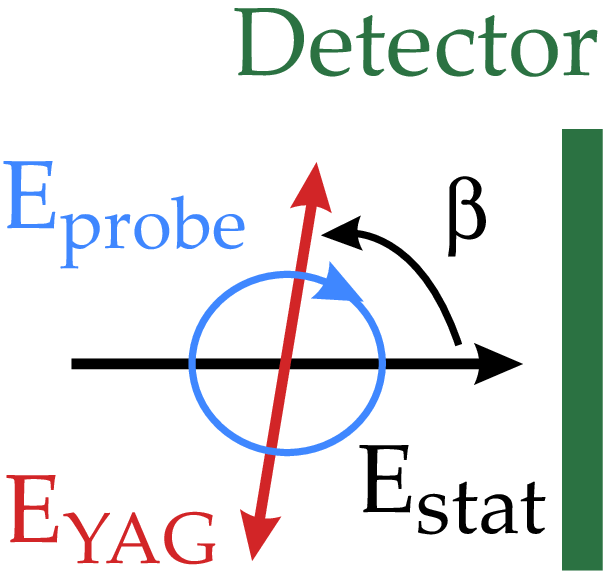}
      \vspace{0.5cm}
   \end{minipage}
   \begin{minipage}{0.79\linewidth}
      \includegraphics[width=\textwidth]{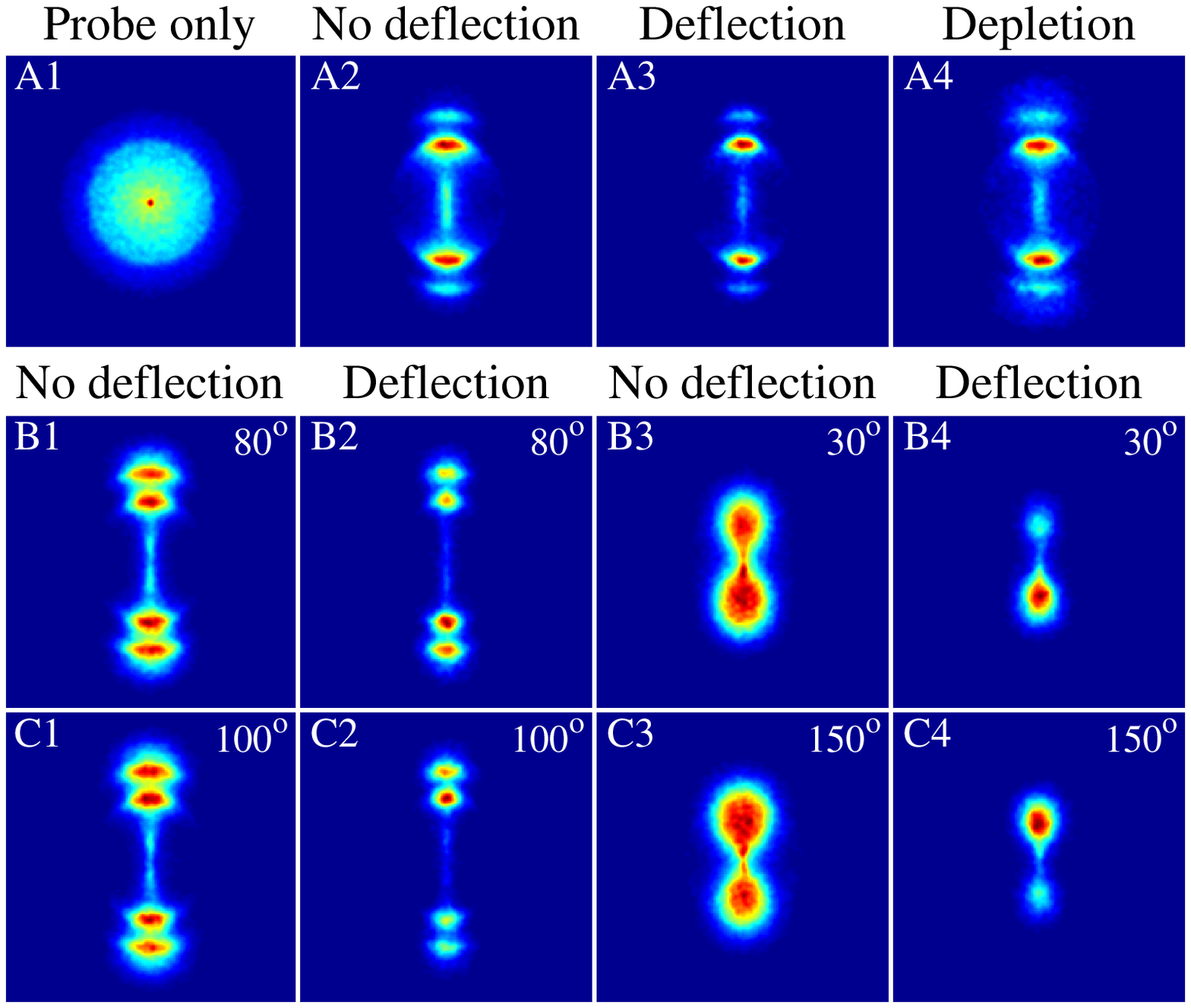}
   \end{minipage}
   \caption{(Color online) I$^+$ ion images, recorded when the probe pulse Coulomb explodes the
      iodobenzene molecules, illustrating alignment and orientation. The polarization state of the
      YAG and the probe pulse with respect to the detector plane and the static field is shown
      schematically at the left. Images labeled ``No deflection'' are recorded at lens position
      $y=0.0$~mm, ``Deflection'' at $y=1.0$~mm and ``Depletion'' at $y=-0.9$~mm, the latter two with
      the deflector at 10~kV (see Fig.~\ref{fig:deflection_IB}). The value of $\beta$ is shown for
      each image in row B and C. The intensity of the YAG pulse is $8\times10^{11}$~W/cm$^2$ and
      $5\times10^{14}$~W/cm$^2$ for the probe pulse. \Es{} = 595~V/cm.}%
   \label{fig:Images}
\end{figure}

We now turn to studying alignment and orientation due to the ac field from the YAG pulse, \Ey{}, and
the static electric field, \Es{}, from the spectrometer electrodes that projects the ions onto the
MCP detector. The basic experimental observables are 2D images of I$^+$ ions recorded when the
molecules are irradiated with both the linearly polarized YAG pulse and the probe pulse. The angular
distribution of the I$^+$ ions provides direct information about the spatial orientation of the C-I
bond axis of the iodobenzene molecules~\cite{larsen:1999:jcp,kumarappan:2006:jcp}. Examples of
images are displayed in Fig.~\ref{fig:Images}.

Image~A2 is recorded with the YAG pulse linearly polarized parallel to the detector plane (vertical
in Fig.~\ref{fig:Images}), corresponding to an angle between \Es{} and \Ey{} of $\beta=90\,\degree$.
In these alignment experiments (row~A of Fig.~\ref{fig:Images}) the Coulomb explosion laser is
linearly polarized perpendicular to the detector plane. The I$^+$ ions appear as angularly narrow
rings. The innermost (and brightest) ring results from I$^+$ ions when iodobenzene is doubly ionized
by the probe pulse and fragments into an I$^+$~+~C$_6$H$_5$$^+$ ion pair, whereas the outermost ring
results from I$^+$ ions formed from triple ionization and fragmentation into an
I$^+$~+~C$_6$H$_5$$^{2+}$ ion pair~\cite{larsen:1999:jcp}. The pronounced angular confinement,
quantified by $\langle\cos^2\theta_{2D}\rangle=0.93$ (calculated from the outer ring, where
$\theta_{2D}$ is the angle between the projection of the I$^+$ recoil velocity on the detector plane
and \Ey{}), shows that the C-I axis of the iodobenzene molecules is strongly aligned along \Ey{}.
For comparison, an image without the YAG pulse is shown (A1). As expected it is circularly symmetric
and $\langle\cos^2\theta_{2D}\rangle=0.51$. These observations are fully consistent with previous
experiments on adiabatic alignment of iodobenzene~\cite{kumarappan:2006:jcp}. When the experiment is
conducted on the most deflected molecules (image~A3) the angular confinement sharpens even further
and $\langle\cos^2\theta_{2D}\rangle$ increases to 0.96~\footnote{When \Es\ is lowered to 297~V/cm
   the angular resolution of the experiment increases and $\langle\cos^2\theta_{2D}\rangle$ values
   up to 0.97 are obtained. We note that $\langle\cos^2\theta_{2D}\rangle=1$ would correspond to the
   quantum mechanically unfeasible situation of perfectly 1D aligned molecules.}. By contrast, when
the experiment is conducted on the least deflected molecules (image~A4), corresponding to the
molecules in the highest rotational states, the alignment weakens and
$\langle\cos^2\theta_{2D}\rangle$ decreases to 0.90.

When the polarization of the YAG pulse is rotated away from the detector plane, the up/down symmetry
characterizing the images in row~A is broken. This is illustrated by the images in row~B and C. In
these experiments, the Coulomb explosion laser is circularly polarized~\footnote{The circular
   polarization of the Coulomb explosion laser avoids changes of the relative polarizations of the
   two lasers, and changes of the angle between the Coulomb explosion laser and the detector plane.
   These changes would artificially affect the obtained images.}. For images with $\beta<90\degree$
(row~B) more I$^+$ ions are detected in the lower part, whereas for $\beta>90\degree$ (row C) more
I$^+$ ions are detected in the upper part. The asymmetry becomes more pronounced as the YAG
polarization is rotated closer to the axis of the static field (compare, for instance images~B1 and
B3, or B2 and B4). We interpret these observations as orientation due to the combined effect of the
YAG laser field and the static electric extraction field~\cite{friedrich:1999:jcp,
   friedrich:1999:jpca}.
The orientation is expected to place the I-end of the molecules towards the repeller plate (see
Fig.~\ref{fig:setup}), where the electrical potential is highest. For iodobenzene all states are
high-field seeking. Therefore, the permanent dipole moment of the molecule, which is directed along
the C-I axis from I (``negative end'') towards the phenyl group (``positive end''), is oriented
parallel to the electric field. Thus, for $\beta>90\,\degree$ the I$^+$ ions are expected to
preferentially be ejected upwards, and for $\beta<90\,\degree$ they will be ejected downwards. This
is in agreement with the up/down asymmetry on the images.

The significant improvement of the up/down asymmetry, or equivalently the orientation, obtained by
using the most deflected molecules is clear by comparing image B1 with B2 (or C1 with C2) and B3
with B4 (or C3 with C4). To quantify the degree of orientation, we determine for each image the
number of I$^+$ ions, N(I$^+$)$_\text{up}$, in the upper part of the I$^+$~+~C$_6$H$_5$$^{+}$ and
I$^+$~+~C$_6$H$_5$$^{2+}$ channels (i.~e., ions detected in the upper half of the images) as well as
the total number of ions,
$\text{N}(\text{I}^+)_\text{total}\left(=\text{N}(\text{I}^+)_\text{up}+\text{N}(\text{I}^+)_\text{down}\right)$.
\begin{figure}
   \includegraphics[width=\linewidth]{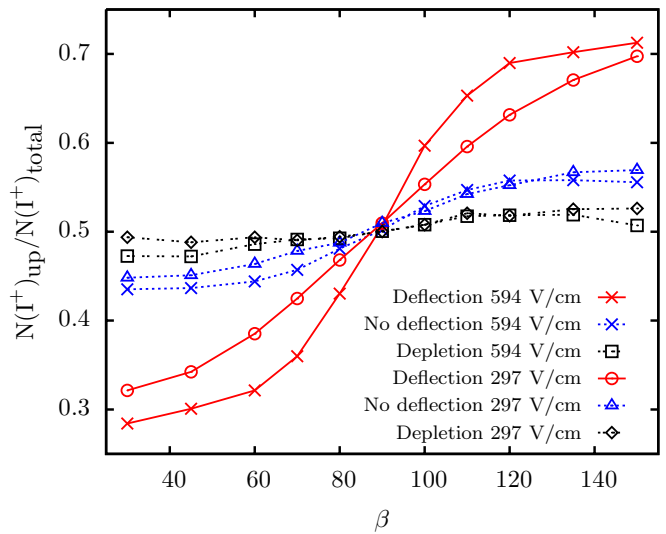}
   \caption{(Color online) Orientation, represented by N(I$^+$)$_\text{up}$/N(I$^+$)$_\text{total}$,
      as a function of $\beta$; see text for details.}
   \label{fig:orientationupdown}
\end{figure}
In Fig.~\ref{fig:orientationupdown} the ratio N(I$^+$)$_\text{up}$/N(I$^+$)$_\text{total}$ is
plotted as a function of $\beta$ for deflected and undeflected molecules and for two different
extraction fields. The difference between the data for the most deflected molecules and the data
obtained with the deflector turned off is striking and shows the advantage of selecting the lowest
lying rotational states for strongly increasing the degree of orientation. From the data for
deflected molecules it is also obvious that increasing the static electric field by a factor of 2
leads to better orientation. In addition, Fig.~\ref{fig:orientationupdown} shows the results
obtained for the least deflected molecules in the depleted region ($y=-0.9$~mm). The degree of
orientation is significantly reduced compared to that obtained without deflection.

Deflection of cold molecular beams enables the selection and the spatial separation of the most
polar quantum states, i.~e., the lowest lying rotational states, for a wide range of molecules, from
diatomics to large biomolecules. In many cases it may even be possible to completely isolate the
rotational ground state. Although the deflection of iodobenzene seeded in helium, treated here, is
already sufficiently selective that laser induced alignment and orientation is significantly
improved, we point out that more stringent state selection is possible by a) lowering the speed of
the molecular beam, for instance by using Ne rather than He as the carrier gas, and, b) increasing
the deflection field gradient or the length of the deflector. Generally, deflection will be more
pronounced for lighter molecules and for more polar molecules.

Adiabatic alignment and orientation, discussed in this work, is one example where state selection by
deflection is highly advantageous. Pronounced effects are also expected in nonadiabatic
alignment~\cite{Holmegaard:pra:2007} and orientation~\cite{cai:2001:prl}, both, in terms of
achieving strong field-free alignment and orientation, as well as in revealing quantum effects
usually obscured by the averaging over many rotational states in a typical molecular ensemble.
Moreover, the deflection spatially separates the selected molecular states from the background of
atomic carrier gas, which could be useful for isolating a molecular signal in high-harmonic
generation and attosecond experiments. Finally, the ability to disperse molecular beams by
inhomogeneous electric fields is not limited to rotational state selection but naturally
differentiates between individual structural isomers of molecules~\cite{Filsinger:prl:2008}. This is
expected to become of significant importance not just in spectroscopic studies but also in emerging
structural and dynamical studies employing pulses from coherent X-ray sources.

\begin{acknowledgments}
   We thank Henrik Haak for expert technical support. This work is further supported by the
   Carlsberg Foundation, the Lundbeck Foundation, the Danish Natural Science Research Council, and
   the Deutsche Forschungsgemeinschaft within the priority program 1116.
\end{acknowledgments}

\bibliography{ref}

\end{document}